\documentclass[a4paper]{appolb}
\usepackage{amsmath}
\usepackage{graphicx}
\usepackage{physics}
\usepackage{subcaption}
\usepackage[compat=1.1.0]{tikz-feynman}

\newcommand{\savg}[1]{\overline{\left|#1\right|}^2}
\newcommand{\LP}{\mathrm{LP}}
\newcommand{\NLP}{\mathrm{NLP}}
\newcommand{\Int}{\mathrm{int}}
\newcommand{\ext}{\mathrm{ext}}

\title{The emission of soft-photons and the LBK theorem, revisited
    \thanks{
        Presented by R. Balsach at Matter To The Deepest Recent Developments
        In Physics Of Fundamental Interactions XLV International Conference
        of Theoretical Physics, Ustroń, Poland, 17-22 September.
    }%
}
\author{
    Roger Balsach, Anna Kulesza
    \address{
        Institut f\"{u}r Theoretische Physik,
        Westf\"{a}lische Wilhelms-Universit\"{a}t M\"{u}nster,
        Wilhelm-Klemm-Stra\ss e 9, D-48149 M\"{u}nster, Germany
    }
    \\[3mm]
    {Domenico Bonocore 
        \address{
            Technical University of Munich, TUM School of Natural Sciences,\\
            Physics Department T31,
            James-Franck-Stra\ss e 1, D-85748,
            Garching, Germany
        }
    }
}

\begin{document}
\maketitle
\begin{abstract}
    Predictions for processes involving soft photons,
    up to next-to-leading power (NLP) in the photon energy,
    can be obtained using the Low-Burnett-Kroll (LBK) theorem.
    The consistency of the theorem has been a recent topic of investigation
    since it is traditionally formulated in terms of a non-radiative
    amplitude, which is evaluated with unphysical momenta.
    We address such questions and propose a formulation of the
    LBK theorem which relies on the evaluation of the non-radiative
    amplitude with on-shell, physical momenta.
    We use this form to numerically study the impact
    of NLP contributions to cross-sections for $pp$ and
    $e^-e^+$ processes involving soft-photon emission.
\end{abstract}

\section{Soft-photon anomaly}

The theoretical framework of radiation in the low-energy (i.e. soft) limit is
based on soft theorems, which enable the computation of
radiative processes solely from
the knowledge of the non-radiative amplitude and the external momenta.
In QED, this universal factorization persists at next-to-leading power (NLP) in
the soft photon energy, as derived long ago by Low, Burnett and Kroll
(LBK) \cite{Low:1958sn, Burnett:1967km}.
Specifically, for an unpolarized cross-section and keeping only the
leading-power (LP) term in the soft expansion,
the soft theorem relates the radiative amplitude ${\cal A}(p,k)$
and the non-radiative amplitude ${\cal H}(p)$ via
\begin{equation} \label{eq:LP}
    \savg{{\cal A}(p, k)}
    = - \left(
        \sum_{ij=1}^{n} \eta_i \eta_j Q_i Q_j
        \frac{p_i \cdot p_j}{(p_i \cdot k) (p_j \cdot k)}
    \right) \savg{{\cal H}(p)}~,
\end{equation}
where $\eta_i$ is $+1$($-1$) for incoming(outgoing) particles.

However,
the data gathered in several hadronic experiments
\cite{DELPHI:2005yew, DELPHI:2010cit, Wong:2014ila} shows a disagreement with
the above formula,
with an excess of photons that ranges between
4 and 8 times the theoretical predictions.
Furthermore, there are plans to upgrade the ALICE detector
\cite{Adamova:2019vkf, ALICE:2022wwr}
with the aim of measuring ultra-soft photons. In light of this long-standing
puzzle and proposed future measurements, further theoretical studies of
soft-photon
emissions are thus necessary. In particular,
it is interesting to estimate the impact of NLP corrections to eq.
\eqref{eq:LP}, as given by the LBK theorem.

\section{Soft-photon emission via the LBK theorem}

\begin{figure}[tbh]
    \centering
    \begin{subfigure}[b]{0.3\textwidth}
        \centering
        \[
        \vcenter{\hbox{\begin{tikzpicture}
            \begin{feynman}
                \vertex[blob, minimum size=0.75cm] (m) at ( 0, 0) {\(H\)};
                \vertex (a) at (1.2,-0.71);
                \vertex (gi) at (-0.53,0.53);
                \vertex (gf) at (0.71,0.9);
                \vertex (a1) at (0.71,-0.71);
                \vertex (a2) at (0.924,-0.383);
                \vertex (a3) at (1,0);
                \vertex (a4) at (0.924,0.383);
                \vertex (a5) at (0.71,0.71);
                \vertex (b) at (1.2,0.71);
                \vertex (c) at (-1.2,0.71);
                \vertex (b1) at (-0.71,0.71);
                \vertex (b2) at (-0.924,0.383);
                \vertex (b3) at (-1,0);
                \vertex (b4) at (-0.924,-0.383);
                \vertex (b5) at (-0.71,-0.71);
                \vertex (d) at (-1.2,-0.71);
                \vertex (e) at (0, -0.32);
                \diagram* {
                    (a1) -- (m) -- (gi) -- (b1),
                    (a2) -- (m) -- (b2),
                    (a3) -- (m) -- (b3),
                    (a4) -- (m) -- (b4),
                    (a5) -- (m) -- (b5),
                    (gi) --[white, photon,
                    momentum={[arrow distance=1.2mm, arrow style=white] \(k\)}
                    ] (gf),
                };
            \end{feynman}
            \draw[decoration={brace}, decorate]
            (d.south west)-- (c.north west)node[pos=0.5, left] {  \(N\)};
            \draw[decoration={brace}, decorate]
            (b.north east)-- (a.south east)node[pos=0.5, right] {  \(M\)};
        \end{tikzpicture}}}
        \]
        \caption{}
    \end{subfigure}%
    \begin{subfigure}[b]{0.3\textwidth}
        \centering
        \[
        \vcenter{\hbox{\begin{tikzpicture}
            \begin{feynman}
                \vertex[blob, minimum size=0.75cm] (m) at ( 0, 0) {\(H\)};
                \vertex (a) at (1.2,-0.71);
                \vertex (gi) at (-0.53,0.53);
                \vertex (gf) at (0.71,0.9);
                \vertex (a1) at (0.71,-0.71);
                \vertex (a2) at (0.924,-0.383);
                \vertex (a3) at (1,0);
                \vertex (a4) at (0.924,0.383);
                \vertex (a5) at (0.71,0.71);
                \vertex (b) at (1.2,0.71);
                \vertex (c) at (-1.2,0.71);
                \vertex (b1) at (-0.71,0.71);
                \vertex (b2) at (-0.924,0.383);
                \vertex (b3) at (-1,0);
                \vertex (b4) at (-0.924,-0.383);
                \vertex (b5) at (-0.71,-0.71);
                \vertex (d) at (-1.2,-0.71);
                \vertex (e) at (0, -0.32);
                \diagram* {
                    (a1) -- (m) -- (gi) -- (b1),
                    (a2) -- (m) -- (b2),
                    (a3) -- (m) -- (b3),
                    (a4) -- (m) -- (b4),
                    (a5) -- (m) -- (b5),
                    (gi) --[photon, momentum={[arrow distance=1.2mm] \(k\)}]
                    (gf),
                };
            \end{feynman}
            \draw[decoration={brace}, decorate]
            (d.south west)-- (c.north west)node[pos=0.5, left] {  \(N\)};
            \draw[decoration={brace}, decorate]
            (b.north east)-- (a.south east)node[pos=0.5, right] {  \(M\)};
        \end{tikzpicture}}}
        \]
        \caption{}
    \end{subfigure}
    \begin{subfigure}[b]{0.3\textwidth}
        \centering
        \[
        \vcenter{\hbox{\begin{tikzpicture}
            \begin{feynman}
                \vertex[blob, minimum size=0.75cm] (m) at ( 0, 0) {\(H\)};
                \vertex (a) at (1.2,-0.71);
                \vertex (gi) at (-0,0.38);
                \vertex (gf) at (0.71,1.1);
                \vertex (a1) at (0.71,-0.71);
                \vertex (a2) at (0.924,-0.383);
                \vertex (a3) at (1,0);
                \vertex (a4) at (0.924,0.383);
                \vertex (a5) at (0.71,0.71);
                \vertex (b) at (1.2,0.71);
                \vertex (c) at (-1.2,0.71);
                \vertex (b1) at (-0.71,0.71);
                \vertex (b2) at (-0.924,0.383);
                \vertex (b3) at (-1,0);
                \vertex (b4) at (-0.924,-0.383);
                \vertex (b5) at (-0.71,-0.71);
                \vertex (d) at (-1.2,-0.71);
                \vertex (e) at (0, 1.4);
                \diagram* {
                    (a1) -- (m) -- (b1),
                    (a2) -- (m) -- (b2),
                    (a3) -- (m) -- (b3),
                    (a4) -- (m) -- (b4),
                    (a5) -- (m) -- (b5),
                    (gi) --[photon,
                            momentum={[arrow distance=1.2mm] \(k\)}] (gf),
                };
            \end{feynman}
            \draw[decoration={brace}, decorate]
            (d.south west)-- (c.north west)node[pos=0.5, left] {\(N\)};
            \draw[decoration={brace}, decorate]
            (b.north east)-- (a.south east)node[pos=0.5, right] {\(M\)};
        \end{tikzpicture}}}
        \]
        \caption{}
    \end{subfigure}
    \caption{
        Diagram (a) corresponds to the amplitude for a general non-radiative
        process,
        with $N$ initial particles and $M$ final ones.
        Diagram (b) corresponds to the photon radiation from an external line,
        and diagram (c) corresponds to the photon radiation from internal lines.
    }
    \label{fig:NMgamma}
\end{figure}
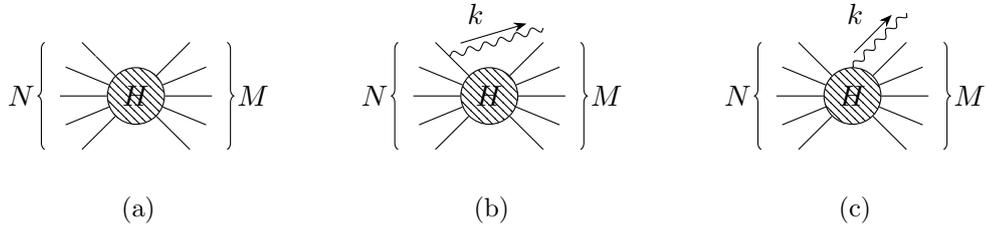

The diagrammatic derivation of the LBK theorem consists of discriminating
the radiative amplitudes $\mathcal{A}^\mu_\ext$ and $\mathcal{A}^\mu_\Int$,
corresponding to radiation coming from the external lines
and internal lines, respectively,
as shown in Fig. \ref{fig:NMgamma}(b) and \ref{fig:NMgamma}(c).
While the contribution from the former can be straightforwardly derived,
internal radiation can be computed using gauge invariance, which yields
\begin{equation}
    \mathcal{A}
    = \varepsilon_\mu \left(\mathcal{A}^\mu_\ext + \mathcal{A}^\mu_\Int\right)
    \Longrightarrow
    k_\mu \left(\mathcal{A}^\mu_\ext + \mathcal{A}^\mu_\Int\right)
    = 0~.
\end{equation}
By exploiting this property, the LBK
theorem at NLP can be easily derived and reads
\begin{align}\label{eq:NLP}
    \savg{\mathcal{A}(p, k)}_{\LP + \NLP}
    =& - \sum_{i, j} \frac{(\eta_i Q_i p_i) \cdot (\eta_j Q_j p_j)}{
        (p_i \cdot k)(p_j \cdot k)
    } \savg{\mathcal{H}(p)} \notag \\&
    - \sum_{i, j} \eta_i Q_i Q_j \frac{p_{i \mu}}{
        (p_i \cdot k)
    }
    G_j^{\mu\nu} \pdv{p_j^\nu} \savg{\mathcal{H}(p)}~,
\end{align}
where
$
    G^{\mu\nu}_j
    = g^{\mu\nu} - \frac{p_j^{\mu}k^{\nu}}{p_j \cdot k}
$.
To derive eq. \eqref{eq:NLP}, the amplitude is expanded in $k$
while considering the other hard momenta $p$ independent of $k$.
Although this step is incompatible with the conservation of four-momentum,
which
can be stated as
\begin{equation}
    \sum_i \eta_i p_i = k~,
\end{equation}
it is mathematically valid.
However,
the consequence is that the non-radiative amplitude
in the right-hand side of eq.
\eqref{eq:NLP} is evaluated using the momenta $p$,
which are unphysical for this process, because $\sum \eta_i p_i \neq 0$.
This might seem problematic
because an amplitude is intrinsically defined for physical momenta,
and it is not uniquely defined for unphysical momenta.
Therefore, the value of $\mathcal{H}(p)$ is ambiguous,
which translates into an ambiguity on $\mathcal{A}(p, k)$
and thus appears to invalidate eq. \eqref{eq:NLP}.
The argument, however, is not entirely correct,
as shown in \cite{Balsach:2023ema}.
Indeed, although an ambiguity is present, it only affects the NNLP terms.
More precisely, if we substitute
$\savg{\mathcal{H}} \to \savg{\mathcal{H}} + \Delta$ in eq. \eqref{eq:NLP}, the
radiative amplitude changes according to
\begin{equation}\label{eq:ambiguity}
    \savg{\mathcal{A}}
    \to \savg{\mathcal{A}}
    + \sum_{i, j} \frac{(\eta_i Q_i p_i) \cdot (\eta_j Q_j p_j)}{
        (p_i \cdot k)(p_j \cdot k)
    } \left[
        1
        + \eta_j \frac{(p_j \cdot k) p_{i \mu}}{p_i \cdot p_j}
        G_j^{\mu\nu} \pdv{p_j^\nu}
    \right] \Delta(p)~.
\end{equation}
However, taking into account that
the function $\Delta(p)$ must vanish when $\sum_i p_i = 0$,
one can see that the ambiguity on
$\savg{\mathcal{A}}$ is in fact of order $\order{1}$ since
\begin{equation*}
    \delta \savg{\mathcal{A}}
    = \sum_{i, j} \frac{(\eta_i Q_i p_i) \cdot (\eta_j Q_j p_j)}{
        (p_i \cdot k)(p_j \cdot k)
    } \left[
        1
        + \eta_j \frac{(p_j \cdot k) p_{i \mu}}{p_i \cdot p_j}
        G_j^{\mu\nu} \pdv{p_j^\nu}
    \right] \Delta(p)
    = \order{1}~.
\end{equation*}
The ambiguity is thus an NNLP effect.
Therefore, the LBK theorem in the form shown in eq.
\eqref{eq:NLP} is consistent and can provide reliable results.
Furthermore, because of the ambiguity in $\mathcal{H}$,
eq. \eqref{eq:NLP} results in an entire family of equivalent formulations.
Nevertheless, from the perspective of numerical implementations,
it would be desirable to have a formulation that fulfils momentum conservation.

\section{Shifted kinematics}

There are multiple ways of restoring momentum conservation
in eq. \eqref{eq:NLP} (see e.g. the discussion in \cite{Burnett:1967km} or,
more recently, in \cite{Lebiedowicz:2021byo, Lebiedowicz:2023mlz}).
The approach we present here is based on the
work described in \cite{DelDuca:2017twk, Bonocore:2021cbv}.
There,
it was proved that the LBK theorem can be reformulated
by shifting the momenta of the non-radiative amplitude. In this
way,
not only the conservation of four-momentum is restored,
but also the derivatives of the non-radiative amplitude are removed, to get
\begin{equation} \label{eq:LBKshifts}
    \savg{\mathcal{A}(p, k)}
    = - \left(
        \sum_{ij=1}^{n} \frac{
            (\eta_i Q_i p_i) \cdot (\eta_j Q_j p_j)
        }{(p_i \cdot k) (p_j \cdot k)}
    \right) \savg{\mathcal{H}(p + \delta p)}~,
\end{equation}
where
\begin{equation}\label{eq:off-shell}
    \delta p_i^\mu
    = Q_i \left(
        \sum_{k, l} \eta_k \eta_l Q_k Q_l
        \frac{p_k \cdot p_l}{(p_k \cdot k)(p_l \cdot k)}
    \right)^{-1}
    \sum_{j} \left(\frac{\eta_j Q_j p_{j \nu}}{k \cdot p_j}\right)
    G_i^{\nu\mu}~.
\end{equation}
\begin{figure}[htb]
    \centerline{%
        \includegraphics[width=0.85\textwidth]{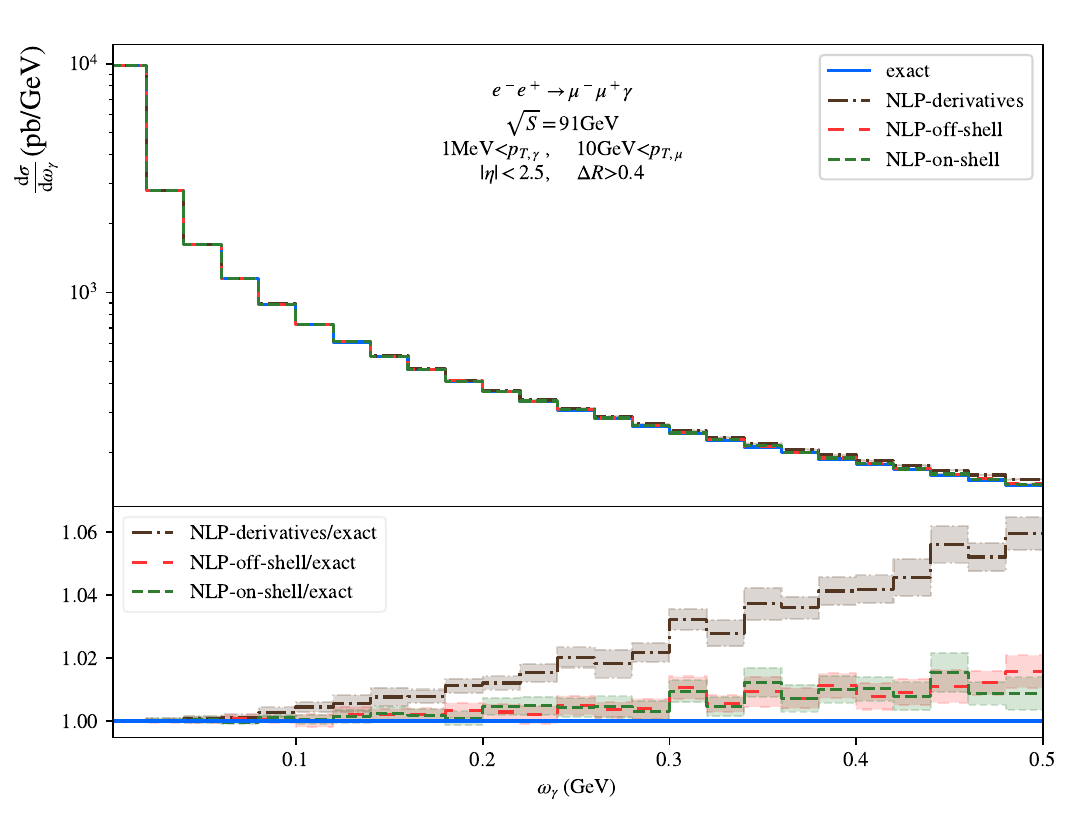}
    }
    \caption{
        Comparison of the soft-photon spectra
        between the different NLP formulations of the LBK theorem,
        normalised to the exact result (i.e. no soft-photon approximation).
    }
    \label{fig:CompNLP}
\end{figure}
\begin{figure}[htb]
    \centerline{%
        \includegraphics[width=0.85\textwidth]{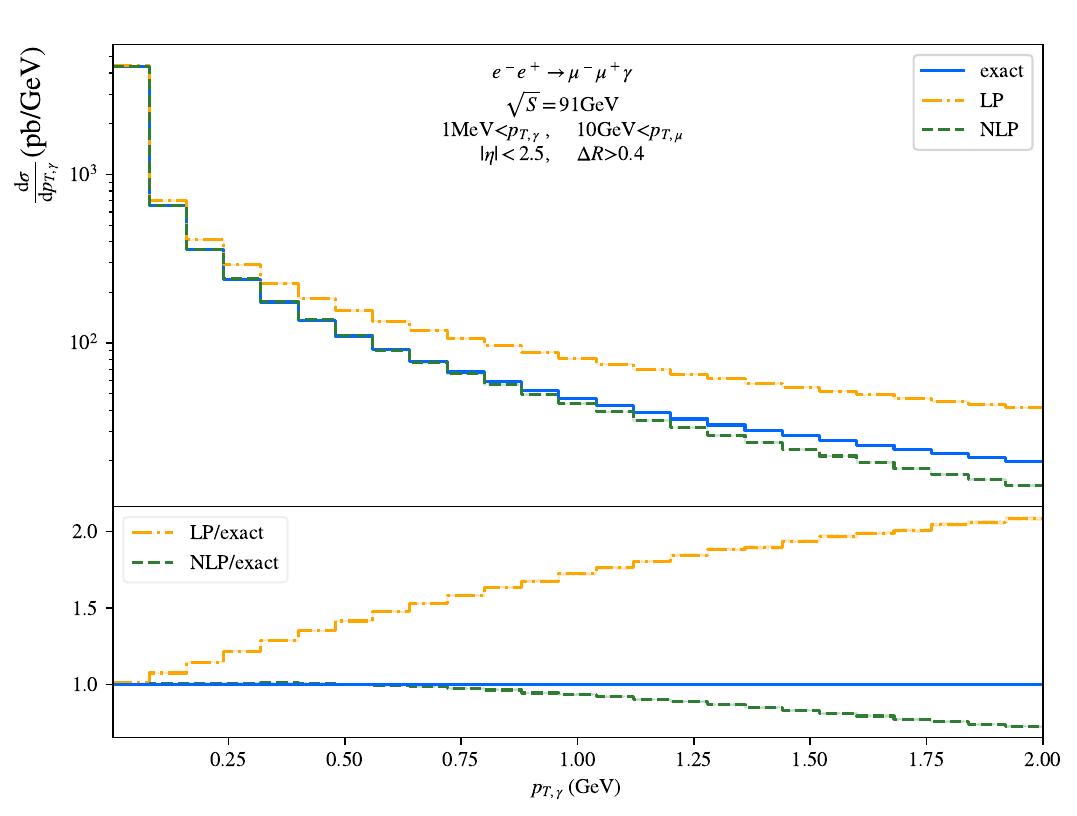}
    }
    \caption{
        Comparison of the $p_T$ distribution calculated at the different
        accuracies in the soft expansion for
        $e^+e^-$ collisions at $\sqrt{S}=91$GeV.
    }
    \label{fig:Compee}
\end{figure}

This formulation simplifies the implementation of the theorem,
making it more suitable for numerical computations that use amplitudes
generated by public tools.
However, it turns out that the shifted momenta $p + \delta p$ in eq.
\eqref{eq:LBKshifts} are not on-shell, which can be problematic in some
applications.
Indeed, one finds
\begin{equation}
    (p + \delta p)^2
    = p^2
    + Q^2_j \left(
        \sum_{k, l} \eta_k \eta_l Q_k Q_l
        \frac{p_k \cdot p_l}{(p_k \cdot k)(p_l \cdot k)}
    \right)^{-1}
    = p^2 + \order{k^2}~.
\end{equation}
To solve this issue, one can
modify the shifts in such a way that they fulfil momentum
conservation and are on-shell to all orders in the soft-photon expansion,
while still keeping eq. \eqref{eq:LBKshifts} valid \cite{Balsach:2023ema}.
These modified shifts are defined as
\begin{align}\label{eq:on-shell}
    \delta p_i^\mu
    = A Q_i \sum_j \frac{\eta_j Q_j}{k \cdot p_j} p_{j\nu} G^{\nu\mu}_i
    + \frac{1}{2} \frac{A^2 Q^2_i {\savg{{\cal S}_{\LP}}}}{p_i \cdot k}
    k^\mu~,
\end{align}
with
\begin{align*}
    A
    = \frac{1}{\chi} \left(
        \sqrt{1 - \frac{2\chi}{\savg{{\cal S}_{\LP}}}}
        - 1
    \right)~,
    \qquad
    \chi = \sum_i \frac{\eta_i Q^2_i}{p_i \cdot k}~,
\end{align*}
\begin{equation}
    \savg{{\cal S}_{\LP}}
    = -\left(
        \sum_{ij=1}^{n}
        \frac{
            (\eta_i Q_ i p_i) \cdot (\eta_j Q_j p_j)
        }{(p_i \cdot k) (p_j \cdot k)}
    \right)~.
\end{equation}
With these new modified shifts,
it is possible to efficiently calculate NLP
soft-photon emissions from arbitrary processes,
and it is possible to use amplitudes numerically generated by public tools.

\section{
    Numerical predictions for $\mu^- \mu^+ \gamma$
    production at $e^-e^+$ and $pp$ collisions
}

Next,
we study the numerical results obtained using
three different versions of the LBK theorem,
i.e. eq. \eqref{eq:NLP}, eq. \eqref{eq:LBKshifts}+\eqref{eq:off-shell} and eq.
\eqref{eq:LBKshifts}+\eqref{eq:on-shell}.
We do so by considering the process $e^- e^+ \to \mu^- \mu^+ \gamma$ and
comparing the results
obtained with the LBK theorem to the exact
results obtained without the soft-photon approximation, as shown in Fig.
\ref{fig:CompNLP}.
One can see that the two formulations with shifts (called ``NLP off-shell'' and
``NLP on-shell'', respectively) seem to work better,
at least for this process,
but no clear difference is visible between the two formulations.

\begin{figure}[htb]
    \centerline{%
        \includegraphics[width=0.85\textwidth]{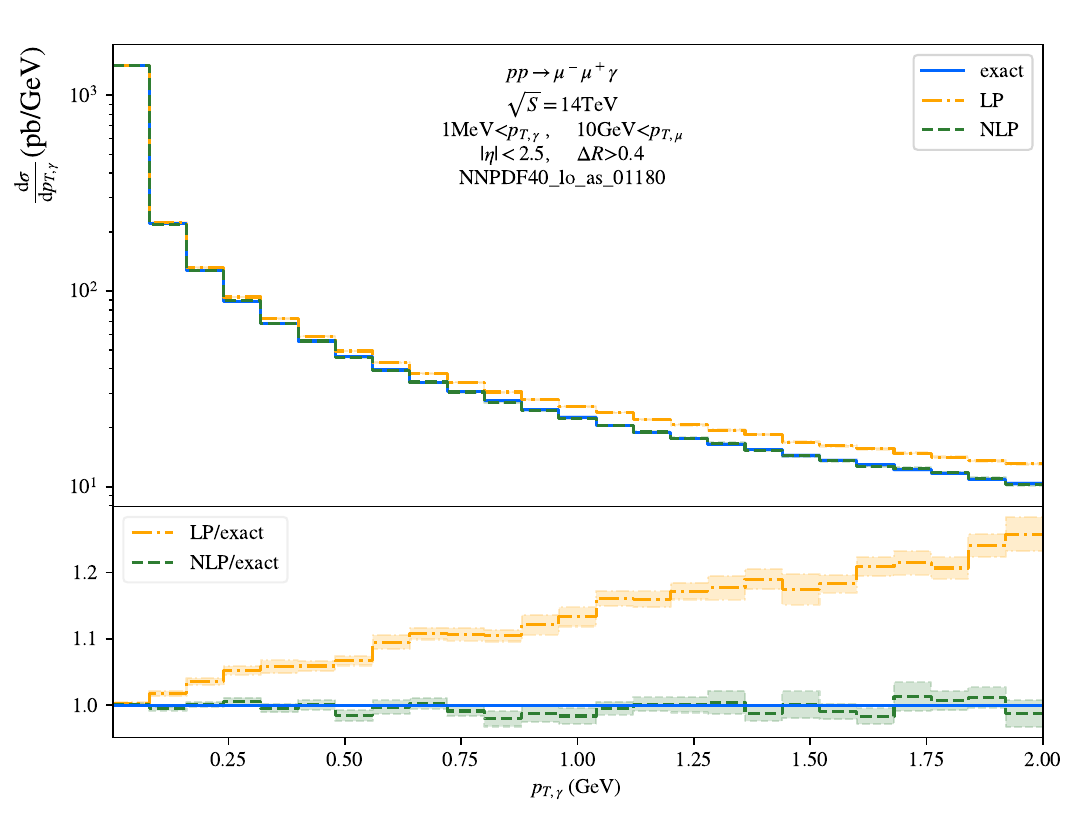}
    }
    \caption{
        Comparison of the $p_T$ distribution calculated at the different
        accuracies in the soft expansion
        for $pp$ collisions at $\sqrt{S}=14$TeV.
    }
    \label{fig:Comppp}
\end{figure}
Since the on-shell shifts of eq. \eqref{eq:on-shell}
enable writing the amplitudes generated numerically,
we use this formulation of the LBK theorem for the remaining analyses shown in
Fig.
\ref{fig:Compee} and \ref{fig:Comppp}.
Specifically, we compare the results for the processes $e^- e^+ \to \mu^-
\mu^+ \gamma$
and $p p \to \mu^- \mu^+ \gamma$, obtained
using no approximation, with the corresponding expansions at LP and NLP.
To do so,
we apply the following kinematic cuts on the external particles:
$p_{T,\gamma} > 1$MeV, $p_{T,\mu} > 1$MeV in the transverse momenta,
$|\eta| < 2.5$ in the absolute pseudo-rapidity
and $\Delta R > 0.4$ for the angular distance between the particles.
For both processes, we see that the NLP terms provide a very good approximation,
with only a few per cent deviation for energies up to 1GeV,
and that the improvement with respect to the LP approximation is notable.

\newpage
\bibliographystyle{utphys}
\bibliography{bib.bib}

\end{document}